\documentclass[preprint]{elsarticle}
\usepackage{graphicx, bm, amsmath, color}
\usepackage{subcaption, enumerate}
\usepackage{multirow}

\journal{Structural Engineering and Mechanics}
\begin{document}

\begin{frontmatter}
\title{Web application for size and topology optimization of trusses and gusset plates}

\author[l1]{S. Krishnamoorthi}
\address[l1]{US Naval Research Laboratory, Washington DC, USA.}
\ead{shankarjeek.krishnamoorthi.ctr.in@nrl.navy.mil}

\author[l2]{Gaurav\corref{cor1}}
\cortext[cor1]{Corresponding Author. Tel.: +91 814 169 4724.}
\address[l2]{Department of Civil Engineering, IIT Gandhinagar, Gujarat, India.}
\ead{gauravs@iitgn.ac.in}

\author[l3]{A. Mandhyan}
\address[l3]{Linde Engineering Pvt. Ltd., Gujarat, India.}
\ead{amar.mandhyan@iitgn.ac.in}

\begin{abstract}
With its ever growing popularity, providing Internet based applications tuned
towards practical applications is on the rise. Advantages such as no external
plugins and additional software, ease of use, updating and maintenance have
increased the popularity of web applications. In this work, a
web-based application has been developed which can  perform size optimization of
truss structure as
a whole as well as topology optimization of individual gusset plate of each
joint based on specified joint displacements and load conditions. This
application is developed using cutting-edge web technologies such as Three.js and
HTML5. The client side boasts of an intuitive interface which in addition to its
modeling capabilities also recommends configurations based on user input, provides
analysis options and finally displays the results. The server side, using a
combination of Scilab and DAKOTA, computes solution and also provides the user
with comparisons of the optimal design with that conforming to Indian Standard
(IS 800-2007). It is a freely available one-stop web-based application to
perform optimal and/or code based design of trusses.
\end{abstract}
\end{frontmatter}
\section{Introduction}
Cloud computing has gained tremendous popularity in recent years with a large
number of applications being designed as Software as a Service (SaaS). SaaS
entails applications that are delivered over the web to end users while core
computations take place on remote server(s). The key advantage of such
applications is that they enable users to perform computationally intensive
tasks using personal machines of modest configurations. In fact, many of these
applications can even be executed on hand held devices.
The ongoing cloud revolution has also shown its footprints in optimization and
structural analysis in the form of web-based tools that have been developed in
the past few years. \citet{eynard2005web} discussed various
benefits obtained by deploying web based technologies in the fields of 
mechanical design and structural analysis.

Optimal design of structures is often the holy grail of engineers and designers. 
A combination of structural (finite element) analysis
software, such as \citet{ANSYS} and optimization packages, such as DAKOTA
\cite{adams2006dakota} is typically employed to achieve optimal designs. Several
commercial finite element packages have recently began providing in-built
optimization modules which can be utilized with analysis routines in a coupled
manner. While such commercial packages are popular among structural designers,
the need for open source and easy to use software cannot be undermined.

SaaS oriented optimization and structural analysis web applications provide an
attractive alternative to equivalent commercial
software packages. In the context of the present study, such applications can be
divided into two broad categories: one concerning structural analysis and
other dealing with optimization. The similarity among the two categories lies in their
client-server model wherein core computations take place at a centralized
location (server) while the so-called front end is provided over the Internet
through various web-based technologies such as HTML, Javascript, PHP, SQL and
DQL, etc.

\citet{peng2002prototype} developed a prototype framework for
Internet oriented collaborative development of a structural analysis program.
They employed OpenSeeS \cite{OpenSeeS}, a freely available nonlinear finite
element package, at the back-end to demonstrate their framework and a typical
web-based interface for user interaction. However, input files for the model
were required to be provided in Tool Command Language (Tcl) \cite{tcl}, a
scripting language understood by OpenSees with no graphical input capabilities.
They did provide a user interface based on Matlab \cite{matlab} using which the
users can download analysis data that can be directly visualized in Matlab.

Later, \citet{Yang2004} developed a web-based platform for computer
simulation of seismic ground response. Their application employed \citet{cyclic1d}, a nonlinear finite element simulation platform for computing
nonlinear seismic ground response of soils. More recently,  \citet{gracia2013integrated} developed an integrated 3D web application for
structural analysis. The work of Gracia and Bayo marked a paradigm shift in the
use of latest web technologies for the purpose of structural analysis. While
earlier efforts required the user to have some level of familiarity with the
underlying computational engine (to be able to write appropriate input files),
Gracia and Bayo alleviated this requirement by providing WebGL-based
\cite{WebGL} interface that allows users to graphically manipulate the
structural model. The core structural analysis module was written in C{}\verb!++! and was
linked to the web interface through PHP. \citet{Hejazi2014}
developed an interactive finite element web application for linear and nonlinear
analysis of reinforced concrete structures subjected to static and dynamic
loading. An exception to the web-based
structural engineering software developed prior to the year 2013 is the one
developed by \citet{Chen2006} which featured a rich Java-based
graphics user interface (GUI) on the client side and a parallel C{}\verb!++!-based finite
element module on the server side.

A web-based post processing tool for finite element analysis was
developed by \citet{Weng2011}. This tool functioned by maintaining a
database of all the finite element input and analysis data, and provided an
improved graphical environment to visualize and post process this data.
Internet-based grid computing environments for structural analysis have also
been developed. Since grid computing is not in the scope of the present work, it will not be
touched upon in detail. Interested readers may refer to the literature
\cite{Alonso2007,Alonso2008,Chen2008,chen2011internet} for more details.

From the optimization standpoint, the studies covered herein involve web based
topology optimization tools. \citet{Tcherniak2001}
developed the first web-based topology optimization tool
\cite{TopOpt} which is, perhaps, the most popular online
topology optimization tool. \citet{Paulino2005} developed a
Java-based topology optimization program with web access while more recently,
\citet{Jorgensen2014} developed a 3D topology optimization
tool for hand held devices and demonstrated its usability on iPad.

Evidently, the available web-based tools for structural analysis and
topology optimization exist in isolation. The present study is the first attempt
in providing a combined application. Moreover, the present application is not
limited just to topology optimization and includes capabilities of DAKOTA
\cite{adams2006dakota} to perform weight and size optimization as well. Key
capabilities of the developed web application include structural analysis of
trusses using matrix method, analysis of plates using finite element method,
design of trusses as per the IS 800:2007 \cite{IS800} (the Indian code of
practice for design of steel structures), optimal design of trusses, and
topology-optimized design
of gusset plates. It is expected that this web application will be
useful for structural engineers and designers as well as students and educators.
Owing to its simplistic and intuitive design, it can be easily utilized to
perform any of the aforementioned tasks and more broadly, can be viewed as a
one-stop solution to 2D truss analysis and design problems. While space trusses
are not yet supported, this capability will be added in the future.
Nevertheless, in many applications, roofs are constructed using a series of 2D
trusses. Such cases can be handled by the application and hence, it can be
utilized for design of supporting trusses of roofs.

We first describe the various aspects of web application and the analysis
capabilities of the web application. Then we answer the question of verification
(``Are we solving the problem right'') by comparing the results of
the web application with analytical solution and finally demonstrate how the web
application can be used to solve typical engineering problems.

\section{Development of Web Application} 
\label{WebApp}
\begin{figure}[!h]
\centering
\includegraphics[scale=0.65]{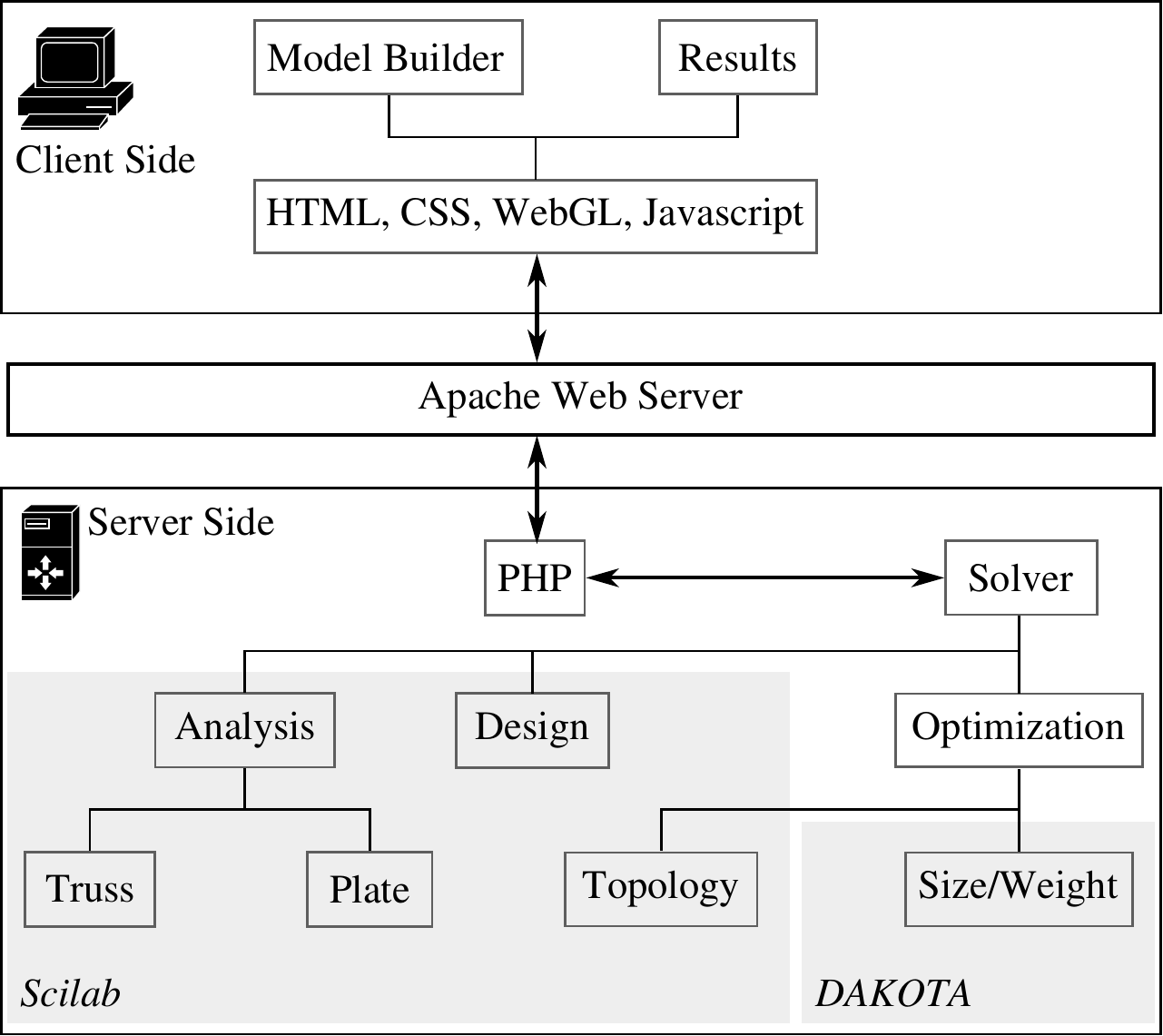}
\caption{Architecture of the Web Application.}
\label{fig:system}
\end{figure}

The web application is hosted on an Apache web server and comprises of several
client and server side components, as detailed in Figure~\ref{fig:system}. The
client side components primarily include a WebGL-based graphical model builder
and a dynamic page that displays analysis results. Server side components
include analysis, design and optimization solvers developed in Scilab
\cite{ScilabEnterprises2012} and DAKOTA \cite{adams2006dakota}. PHP is employed
to manage the flow of data (input/output of parameters and results to/from the
solver). Thus, the user only requires a web browser with a working Internet
connection to use the web application as all the computationally intensive tasks
are carried out on the server side.

\subsection{Input Interface}
\begin{figure}[!h]
\centering
\includegraphics[width=1\textwidth]{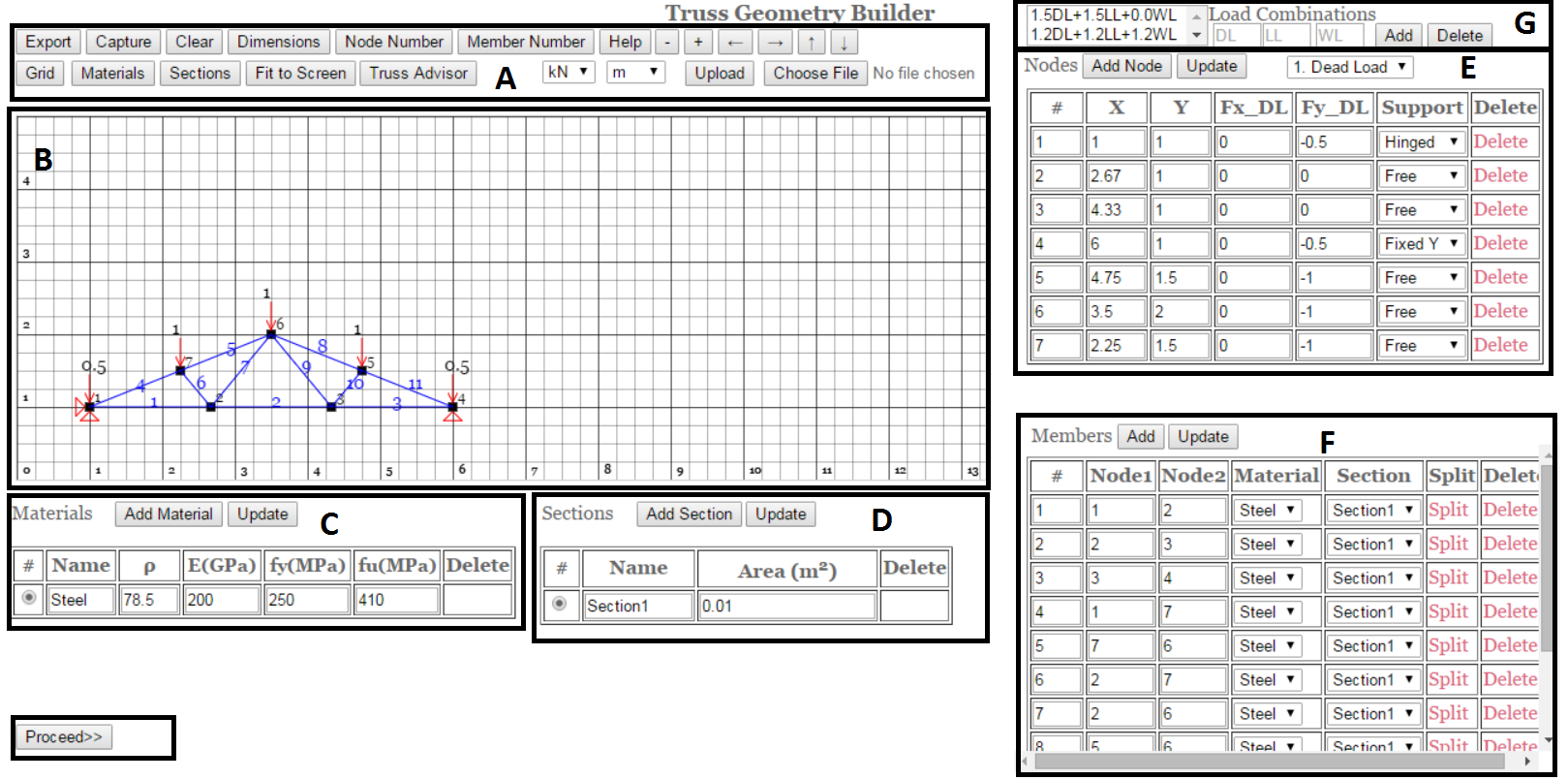}
\caption{Input Interface of the Web Application: Screen 1.}
\label{fig:input}
\end{figure}

The input interface has been designed using latest web technologies such as
Three.js, HTML5, CSS and Javascript and is shown in Figure~\ref{fig:input}. The
input interface comprises of eight main components demarcated by blocks $A$
through $H$ in Figure~\ref{fig:input}. Block $A$ is a set of buttons that provides
all the action items of the application. Block $B$ is the drawing canvas made using
Three.js (an extension of WebGL) that allows the users to construct geometries
of the problems they wish to analyze.

Nodes can be specified by right-clicking the mouse within the canvas. By default, nodes are placed at grid points and the node table in block $E$ is automatically populated with newly added nodes. Finer adjustments of node positions can be made by altering the `X' and `Y' coordinate values in block $E$. Block $E$ can also be utilized to prescribe nodal loads and nodal support conditions. User can specify dead load (DL), live load (LL) and wind load (WL) and any number of their combinations. A combination can be added by specifying corresponding factors for dead load, live load and wind load as shown in Block $G$. Load combination suggested by the web application are 1.5(DL + LL), 1.2(DL + LL + WL) and 1.5WL + 0.9DL.  Presently, three types of support conditions, sufficient to describe trusses, can be specified: roller, hinged, and free; by default, every node is free. Deletion of nodes, if required, can be performed using the `Delete' option provided against respective nodes.

Members or elements can be added by left-clicking on the starting node and
dragging the mouse to the ending node. The member table in block $F$ is automatically
populated with the newly formed members. Just like nodes, members can be deleted
by using the `Delete' link provided against respective members. An additional
feature `Split' is provided that can be used to introduce an additional node at
the mid-point of the member.

Blocks $C$ and $D$ are dynamic tables where users can add materials and 
cross-sectional properties to be used in the structural model. Poisson ratio, $\nu$, Young's
modulus, $E$, yield stress, $f_\mathrm{y}$, and ultimate stress, $f_\mathrm{u}$,
are required to be specified for each material. By default, block $C$ contains
properties corresponding to steel. As for cross-sectional properties, the
area of cross section is required to be provided; the default value for which is
$0.01 \mathrm{m}^2$. All the newly created members are assigned default values
for material and cross-section which can be changed in block $F$.

\begin{figure*}[!h]
\centering
\includegraphics[width=1\textwidth]{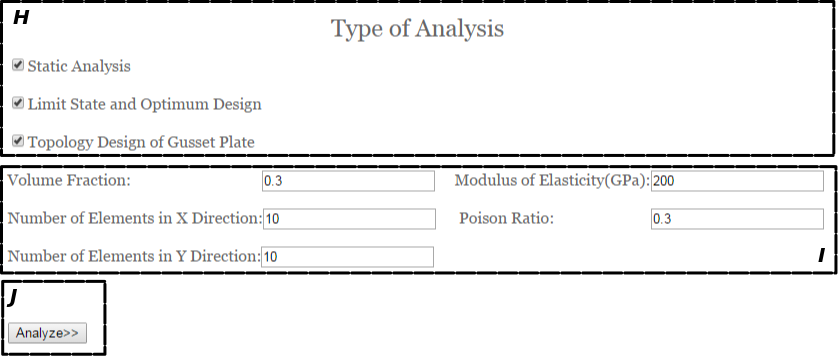}
\caption{Input Interface of the Web Application: Screen 2.}
\label{fig:input_2}
\end{figure*}

Once all the inputs have been specified, user clicks on the `Proceed' button provided in block $G$ that takes the user to the next input page in which the types of analyses/design to be carried out can be specified. The second input page is shown in Figure~\ref{fig:input_2}. Block $H$ shows the types of analyses supported by the web application. The user can select one or more of the analysis/design options. Block $I$ shows parameters relevant to the gusset plates and topology optimization. The input `Volume Fraction' is specified for carrying out the topology optimization of gusset plates. Number of elements in X and Y directions are utilized to specify the size of the finite element mesh to be employed for analysis of plates. Finally, when all the inputs have been specified, the `Analyze' option in block $J$ can be utilized to begin the analysis. Once this button is clicked, all the model data is sent to the server
where the analysis is carried out.

While the input interface has been described with respect to building a truss
model, the interface is similar for modeling of plates. Users get access to some
additional features, such as choosing the type of 2D idealization (plane stress
or plane strain) and specification of thickness in case of plates. Full details are not provided for the sake of brevity.

\subsection{User Friendly Features}
Several user friendly features are provided through the buttons provided in
block $A$.
It is recognized that users may wish to save their work and/or reuse previously
generated models. The `Export' option allows users to save the entire model as a
\texttt{csv} file that can be utilized to work with the saved model at a later
time through the `Choose File', `Upload' and `Show' options. The `Capture'
option allows users to save a screen-shot of the present view of the model while
`Dimensions', `Node Number', `Member Number', `Grid', `Show Materials', and
`Show Sections' are toggle options provided to enable showing/hiding of relevant
visual features. When the `Show Materials' or the `Show Sections' options are
switched on, the members are color coded with respective materials and/or
sections. Further, it is possible to zoom and pan the canvas using the
`+', `-', `$\leftarrow$', `$\rightarrow$', `$\uparrow$', and `$\downarrow$'
options. The option of `Fit to Screen' is also provided to facilitate zooming.

\begin{figure}
\centering
\includegraphics[width=1\textwidth]{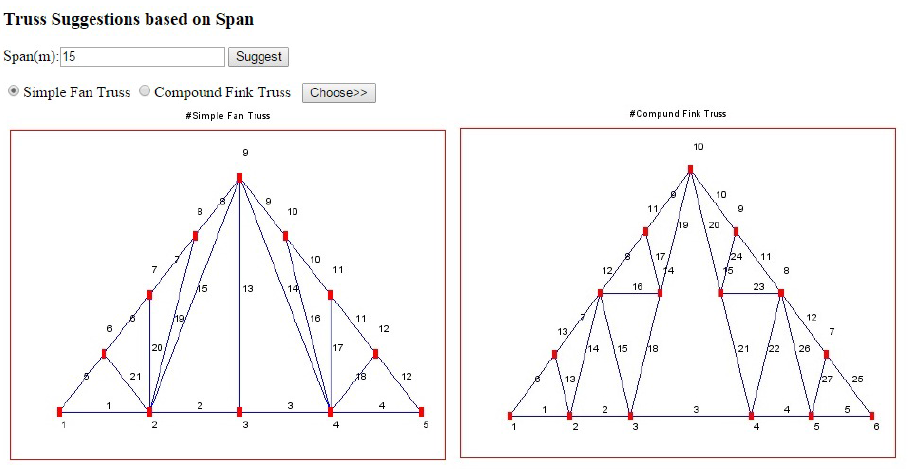}
\caption{Truss Advisor}
\label{fig:advisor}
\end{figure}

In order to make the web application more accessible for persons with limited
knowledge of truss systems, a `Truss Advisor' option is provided. Once utilized,
it leads the user to a help page, shown in Figure~\ref{fig:advisor}. In this page,
the user can specify the total span to be covered by the truss and the web
application shows relevant suggestions. If the user chooses a suggestion from
this page, it automatically gets loaded in the main input section. It is
possible to customize the pre-loaded truss system by making use of tables $C$
through $F$ on the main input page.

\subsection{Core Computational Routines}
The core computational routines correspond to the types of analyses the web
application is capable of performing. First of these is static analysis of
trusses. A matrix analysis based method has been implemented in Scilab to carry
out analysis of trusses subjected to static loads. Through this analysis, the
forces in all members and deflections at all nodes are computed.

Next is limit state design of trusses as per the code of practice for steel
structures published by Bureau of Indian Standards (BIS), IS 800:2007
\cite{IS800}. The design algorithm has been implemented in Scilab and comprises
of a library of $136$ standard angle sections provide in IS 800:2007. The
peripheral members of the truss are designed as using double angle sections while
the interior ones are designed as single angle sections. Connections of members
and gusset plates are assumed to be $4~\mathrm{mm}$ fillet weld connections
in accordance with IS~800:2007. All the checks prescribed by
IS~800:2007 for tension and compression members are enforced by the design
algorithm. Zero force members are assigned section of minimum weight from the
library.

In addition to analysis and limit state design of trusses, the web application
can also address weight/size optimized design of trusses. A
combination of Scilab routines and DAKOTA is utilized to achieve this. DAKOTA
offers a wide range of gradient-based and non-gradient-based optimization
algorithms. In the optimization problem, minimization of the weight of the truss
is posed as the objective function. Member stresses are constrained to remain
within the elastic limit of the material and the slenderness ratios of the members are 
constrained to remain within the limits prescribed by IS~800:2007. The 
\texttt{conmin\_mfd} optimizer of DAKOTA is being utilized which is a constrained 
minimization algorithm \cite{Vanderplaats1973}.

To consummate the optimal design of the truss assembly a complete specification
of gusset plates is required. In order to achieve optimal gusset plate designs,
a topology optimization routine implemented in Scilab is utilized. All the
gusset plates of the structural model are initially assumed to be rectangular
and the member forces are distributed equally within the welded portion (actual
connection of gusset plate and truss member). The weight/volume of the gusset is
optimized following the same algorithm proposed by \citet{sigmund200199}.
Current implementation does not include addition of stress/displacement
constraints. It is assumed that the axis of all the members connected to a gusset
plate intersect at the centroid of the plate and the members do not overlap,
i.e., adjacent members sharing a gusset plate only touch each other at the vertices.
With this assumption and given orientation of each member identifying the
position of each member in a gusset plate is a trivial task.

The member forces are transferred to the gusset plate in two steps. First, all
the nodes of the FE model of the plate that lie within on the weld region are
identified and then, the member forces are applied to these nodes with equal
distribution. From a numerical, essential boundary conditions also need to be
specified to ensure a unique displacement field. Minimal boundary conditions are
chosen to eliminate rigid body modes thereby not influencing the solution
significantly. The centroid is fixed to prevent translations and horizontal
translation for a node below the centroid location is also prevented to
eliminate rigid body rotations.

The initial density of the material is assumed to be constant for all elements
of the FE model of the gusset plate. The solid isotropic material with
penalization (SIMP) approach \cite{Bendsoe1995}, which utilizes a penalizing
factor to ensure continuous design variables are enforced towards black and
white design, is employed to achieve topology optimized shapes of the gusset
plates.

\subsection{Output Interface}
The results are displayed on the client side in the form of tables, downloadable
text files, and images. Forces in each member, design summary
as per IS 800:2007 and the optimization results from DAKOTA are displayed in a
tabular format wherein they can be readily compared. Furthermore, detailed design 
results with a comprehensive step-by-step summary are available for download by
the user in \texttt{txt} format. Topology optimization results are displayed as 
JPEG images which can also be downloaded by the user. More details of the
output interface are discussed in later sections.

\section{Verification Studies} 
\label{Ref}
In applications related to science and engineering there is always a need for
evaluating the reliability of the computer model. Rigorous standards have been
developed in the field of fluids and solid mechanics to answer the question of
``Verification and Validation''. Verification deals with showing
that the developed numerical model shows good convergence and has error bounds
on quantities of interest and validation deals with the correlation between the
model results and what is really observed. Since standard matrix analysis,
finite element analysis for plane stress/plane strain and topology optimization
have been validated by researchers we focus over attention to verification of
the web application. We consider three cases which serve as a means of verification 
for (a) matrix analysis module for solving truss assembly, (b) finite element
module for plane stress/strain analysis, and (c) topology optimization module.

\subsection{Matrix analysis module}
\begin{figure*}[!h]
    \centering
    \begin{subfigure}{0.475\textwidth}
        \includegraphics[width=\textwidth]{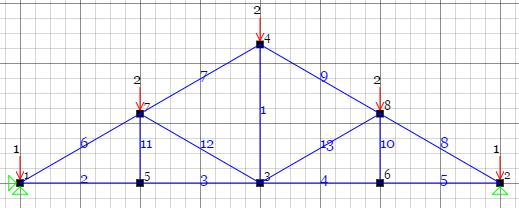}
        \caption{Truss model}
        \label{fig:Verification1}
    \end{subfigure}
    \begin{subfigure}{0.475\textwidth}
        \includegraphics[width=\textwidth]{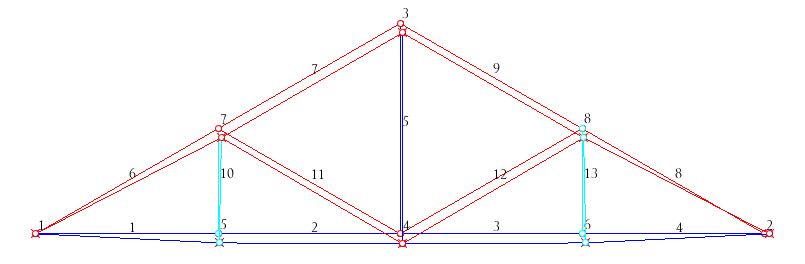}
        \caption{Deflected shape}
        \label{fig:Verification1deformation}
    \end{subfigure}
\caption{Verification of Matrix analysis module of the web application. The undeformed nodal position are represented by filled circles and deformed nodal position by unfilled circles.}
\label{fig:Verification}
\end{figure*}

A simple symmetric truss assembly shown in Figure~\ref{fig:Verification1} was
chosen as the verification problem. The span was taken to be $8$m and the height
was taken as $2.31$m. Loads applied to this assembly were also chosen to be
symmetric.
The deformed shape of the truss assembly as computed by the web application is
shown in Figure~\ref{fig:Verification1deformation} (presently, the deformed
shape is not reported in the web application. This plot was generated for the
purpose of verification). The deformed shape shows a symmetric profile which is
expected from the inherent symmetry in the problem. The individual member force
values computed by the web application show good agreement with the analytical
solution (paper and pen) (Table \ref{tab:Verification1}).

\begin{table}
\caption{Comparison of results between analytical and web application. Positive values denote tension and negative denotes compression.}
\label{tab:Verification1}
\centering
\small
\resizebox{\textwidth}{!}{%
\begin{tabular}{ccccccccccccccc}
\hline \hline
Member &  & 1 & 2 & 3 & 4 & 5 & 6 & 7 & 8 & 9 & 10 & 11 & 12 & 13 \\ \hline
\multirow{2}{*}{Forces (kN)} & Web App & -5.19 & -5.19 & -5.19 & -5.19 & -2 & 5.99 & 3.99 & 5.99 & 3.99 & 0 & 1.99 & 1.99 & 0 \\
 & Analytical & -5.2 & -5.2 & -5.2 & -5.2 & -2 & 6 & 4 & 6 & 4 & 0 & 2 & 2 & 0 \\
Error &  & -0.19\% & -0.19\% & -0.19\% & -0.19\% & -0.00\% & -0.17\% & -0.25\% & -0.17\% & -0.25\% & -0.00\% & -0.50\% & -0.50\% & -0.00\% \\ \hline \hline
\end{tabular}
}
\normalsize
\end{table}

\subsection{Finite element module}\label{FEM}
\begin{figure}[!h]
	\centering
	\includegraphics[width=0.5\textwidth]{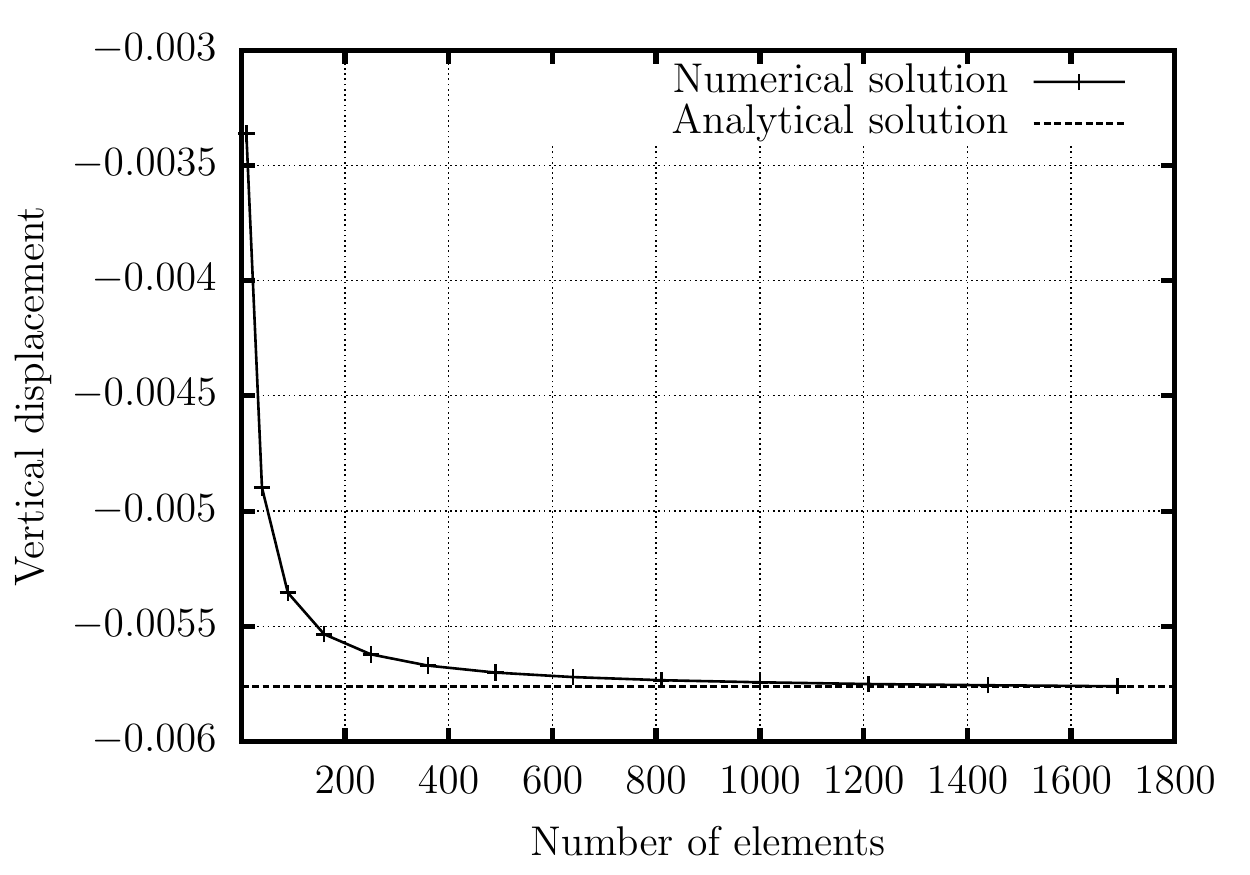}
	\caption{The numerical solution approaches the analytical solution as the number of elements is increased}
	\label{fig:FEconvergence}
\end{figure}

A cantilever beam of length $1.8$m, thickness $0.12$m and depth $0.15$m was solved with a point load of $20$kN applied at the end. The mesh was refined and the tip displacement was compared against analytical solution $(PL^3/(3EI))$ as shown in figure \ref{fig:FEconvergence}. If $n$ is the number of elements through the depth, then $10n$ elements were set along the length. In the finest mesh (1690 elements) the error in values is $0.005\%$.

\subsection{Topology optimization module} \label{TopoPlate}
\begin{figure*}[!htbp]
\centering
\includegraphics[width=0.475\textwidth]{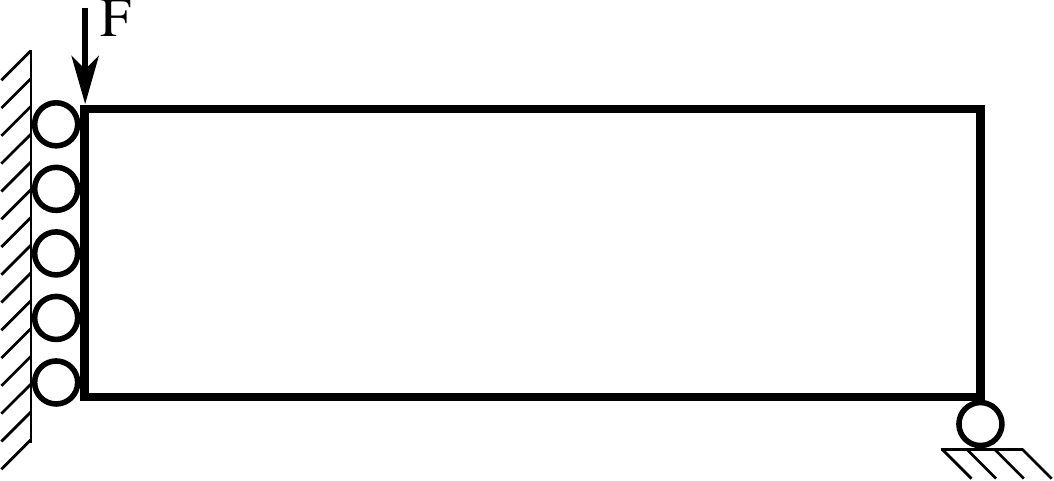} \hfil
\includegraphics[width=0.475\textwidth]{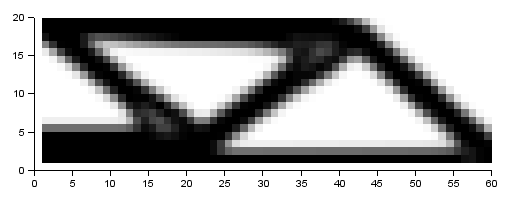}
\caption{The design domain with boundary conditions and loads is shown in the left. The predicted optimal shape is shown in the right.}
\label{fig:MBB}
\end{figure*}

The classical example of minimizing compliance to obtain the optimal
material distribution in the  Messerschmitt-B\"{o}lkow-Blohm beam (MBB) problem
has been chosen as the verification problem for the topology optimization algorithm. The model consists of a plate which is fixed at the bottom left and rests on a roller in the bottom right end. An edge load is applied in the top middle part. In the present study, only half the plate has been modeled and symmetry boundary conditions have been applied to the middle plane of the beam, as shown in
Figure~\ref{fig:MBB}. The geometry, material properties, loads and optimization
parameters were chosen similar to the problem solved by \citet{sigmund200199}. The resulting volume fraction density distribution
obtained by the web application is also shown in figure \ref{fig:MBB}. This
results are good agreement with distribution obtained by \citet{sigmund200199}.

\section{Demonstration of Web Application} 
\label{Demo}
Truss configuration shown in Figure~\ref{fig:input} has been considered for static
analysis, limit state design and optimization of gusset plates. The loads were scaled by a factor of $10$. The results from
limit state design analysis are displayed in Table \ref{tab:Res} where the cross
sections as suggested by IS~800:2007 are shown along with the optimal values of
the corresponding cross sectional areas. The optimized results evidently motivate
the use of the suggested optimal design instead of the classical code-based design.

\begin{table*}[!htbp]
\caption{Limit state analysis results comparing IS-800:2007 code results with optimal values}
\label{tab:Res}
\resizebox{\textwidth}{!}{%
\begin{tabular}{c c c c c c}
\hline\hline\noalign{\smallskip}
Member & Forces (kN) & Length(m) & ISA L(mm) x B(mm) x t(mm) & Area (mm$^2$) & Optimized Area (mm$^2$) \\
\noalign{\smallskip}\hline\noalign{\smallskip}
1 & -37.50 & 1.66 & 2 x ISA 20 x 20 x 4 & 290 & 150.00 \\
2 & -24.99 & 1.66 & 2 x ISA 20 x 20 x 4 & 290 & 100.00 \\
3 & -37.49 & 1.66 & 2 x ISA 20 x 20 x 4 & 290 & 150.00 \\
4 & 40.38 & 1.34 & 2 x ISA 25 x 25 x 5 & 450 & 161.55 \\
5 & 33.64 & 1.34 & 2 x ISA 40 x 25 x 3 & 376 & 134.59 \\
6 & 9.76 & 0.65 & 1 x ISA 20 x 20 x 4 & 145 & 44.35 \\
7 & -9.75 & 1.30 & 1 x ISA 20 x 20 x 4 & 145 & 45.59 \\
8 & 33.64 & 1.34 & 2 x ISA 40 x 25 x 3 & 376 & 134.59 \\
9 & -9.75 & 1.30 & 1 x ISA 20 x 20 x 4 & 145 & 45.59 \\
10 & 9.76 & 0.65 & 1 x ISA 20 x 20 x 4 & 145 & 44.35 \\
11 & 40.38 & 1.34 & 2 x ISA 25 x 25 x 5 & 450 & 161.55 \\
\noalign{\smallskip}\hline\hline                             
\end{tabular}
}
\end{table*}

In addition to this information, a detailed analysis report is generated in text
format, as shown in Appendix~\ref{App:DetailedResult}, which can be downloaded by
the user. Appendix~\ref{App:DetailedResult} shows results reported for one
member under
tension and another under compression and highlights the key differences in
calculations related to the two cases.

\begin{figure*}[!htbp]
    \centering
    \begin{subfigure}{0.3\textwidth}
        \includegraphics[width=\textwidth]{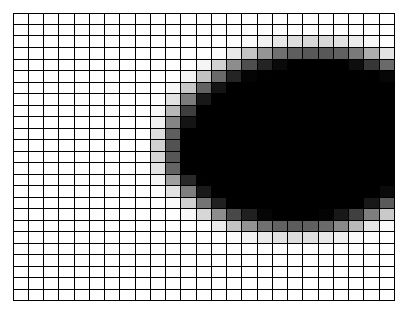}
        \caption{Gusset Plate 1}
        \label{fig:GP1}
    \end{subfigure}
    \begin{subfigure}{0.3\textwidth}
        \includegraphics[width=\textwidth]{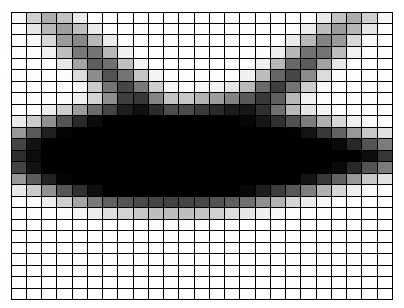}
        \caption{Gusset Plate 2}
        \label{fig:GP2}
    \end{subfigure}
    \begin{subfigure}{0.3\textwidth}
        \includegraphics[width=\textwidth]{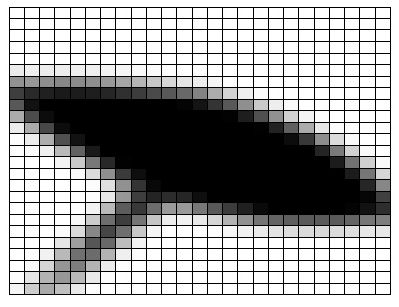}
        \caption{Gusset Plate 5}
        \label{fig:GP5}
    \end{subfigure}

  \begin{subfigure}{0.3\textwidth}
    \includegraphics[width=1\textwidth]{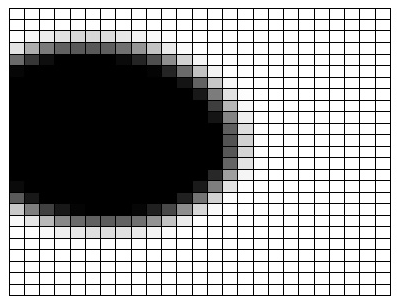}
    \caption{Gusset Plate 4}
    \label{fig:GP4}
  \end{subfigure}%
  \begin{subfigure}{0.3\textwidth}
  \includegraphics[width=1\textwidth]{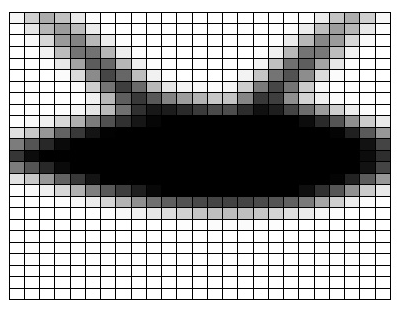}
  \caption{Gusset Plate 3}
  \label{fig:GP3}
  \end{subfigure}
  \begin{subfigure}{0.3\textwidth}
  \includegraphics[width=1\textwidth]{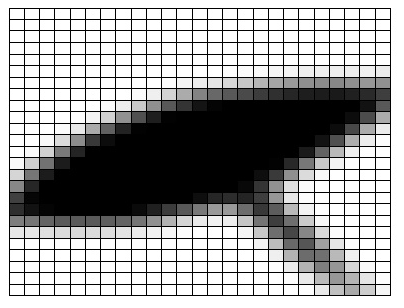}
  \caption{Gusset Plate 7}
  \label{fig:GP7}
  \end{subfigure}%

  \begin{subfigure}{0.3\textwidth}
  \includegraphics[width=1\textwidth]{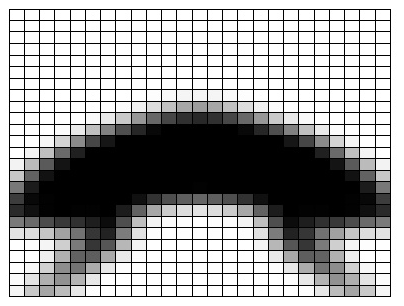}
  \caption{Gusset Plate 6}
  \label{fig:GP6}
  \end{subfigure}
\caption{Optimized shape of all the gusset plates. The number represents the nodal location at which the gusset plate is located. The plates number are positioned to show symmetry in results.}
\label{fig:OptGussetPlates}
\end{figure*}

The web application displays the topology optimization results in the form of
images. These optimal shapes of the gusset plates for the problem under
consideration are shown in Figure~\ref{fig:OptGussetPlates}. The gusset plates
are identified by node numbers mentioned in Figure~\ref{fig:input}. 
Since the chosen truss system is symmetric about its central axis, the optimized
shapes of the gusset plates are also expected to be symmetric in nature. This
can be observed in the shapes of the gusset plates. Gusset plate 6 which is
located on the symmetry line of the truss shows a symmetric shape. The shapes of 
Plates 1-4, 2-3 and 5-7 are mirror images of each other, which is also expected
as these plates are symmetrically located with respect to the symmetry line of
the truss.

\section{Summary} 
\label{Summary}
A web-based tool has been developed which provides a single
interface to analysis, design and topology optimization in
the context of truss structures. The front end is delivered to the end user
through a web browser utilizing the latest web technologies such as Three.js and
HTML5. The core computation routines are executed on the server utilizing a
combination of PHP, Scilab and DAKOTA. The analysis and design results are
routed to the user in the form of tables, images, and downloadable text files.

The developed web application is the first of its kind that combines analysis
and design with optimization and has immense potential both from an educational
and industrial standpoint. The WebGL-based graphical user interface is intuitive
and easy to use for both professionals and those who are new in the field.
Current developments include extension of the web application to 3D and inclusion of constraints in optimization which would allow real time solution of realistic and practical problems. 

\section*{Acknowledgements} 
The authors gratefully acknowledge the financial support provided by Indian
Institute of Technology Gandhinagar that facilitated this work.

\section{References} 
\bibliography{Truss-WebApp-Shankarjee}
\appendix
\section*{Appendix}
\section{Details of steel design of truss (Figure 2) as per IS:800-2007} \label{App:DetailedResult}
\begin{verbatim}
Steel Design of Member No:1 (Tension Member)

Yeild Strength of Steel(fy): 250.00 N/mm$^2$ 
Ultimate Strength of Steel(fu): 410.00 N/mm$^2$ 
Force in the member(F): 3.75 kN 
Length of the member(Lo): 1.67 m 
Providing double angle section
Area of each Section Required to resist the force:8.25mm$^2$  
Assuming Weld Size: 4 mm 
  Strength of Weld= weld_size*0.7*0.462*fu: 530.38 N/mm$^2$  

Section Selected: 20 x 20 x 4 mm 
  Gross Area of Section(Ag): 145.00 mm$^2$  
  Net Area of Section(An): 145.00 mm$^2$  

Strength of Selected Section:-
  Strength due to yeilding of Gross Area=
  (0.91*Ag*fy)/1000=32.99 kN 
  Strength due to rupture of Net Area=
  (0.8*0.8*An*fu)/1000=38.05 kN 
  Strength due to Block Shear:
    Tb1=((0.525*Avg*fy+0.72*Atn*fu)/1000): 56.27 kN 
    Tb2=(0.416*Avn*fu+0.91*Atg*fy)/1000: 60.63 kN 
  Strength due to Block Shear:56.27 kN 

Tensile Strength of Selected Section:32.99 kN > 1.88 kN 

Check For Slenderness Ratio Limits:- 
  Effective Length of the member=Le=(Lo*1.0):1.67 m 
  Minimum Radius of Gyration=Rmin:0.58 mm
  Slenderness Ratio of the member=
  (Le*1000)/Rmin:287.50 < 350 

Provide 2 angle 20 x 20 x 4 with weld size 4 mm all 
along all three edges.
 
Steel Design of Member No:4 (Compression Member)

Yeild Strength of Steel(fy): 250.00 N/mm$^2$ 
Ultimate Strength of Steel(fu): 410.00 N/mm$^2$ 
Force in the member(F): 4.04 kN 
Length of the member(Lo): 1.35 m 
Providing double angle section
Area of each Section Required to resist 
the force: 22.44 mm$^2$  
Section Selected: 20 x 20 x 4 mm 
  Area of Section(Ag): 145.00 mm$^2$  

Check For Section Classification:- 
  Shorter Arm to thickness ratio 
  b/t=5.00 < (15.7*e=15.70)
  Longer Arm to thickness ratio  
  L/t=5.00 < (15.7*e=15.70)
  Sum of Shorter Arm and   Longer Arm to thickness 
  ratio: (L+B)/t=10.00 < (25*e=25.00)
Hence full area of section is effective.

Strength of Selected Section using equations:
  (Design Compressive Strength)
  f_cd=(fy/$\lambda_m$)/sqrt($\phi$+($\phi^2$-$\lambda^2$)))
  $\lambda_{vv}$=((Lo/rmin)/e*sqrt(E$\pi^2$/fy))= 1.19
  $\lambda_{\phi}$=(((L+b)/2*t)/e*sqrt(E$\pi^2$/fy))= 0.06

Assuming Hinged Connection:
  k1=0.7 k2=0.6 k3=5
  $\lambda_{eh}$=sqrt(k1+k2*$\lambda_{vv}^2$+k3*$\lambda_{\phi}^2$)= 1.25
  $\phi$=0.5*[1+$\alpha$($\lambda_{eh}$-0.2)+$\lambda_{eh}^2$]= 1.54
  c=($\phi$+sqrt($\phi^2$-$\lambda_{eh}^2$))= 2.44
  f_cd=(fy/$\gamma_{mo}$)/sqrt($\phi$+($\phi^2$-$\lambda^2$)))=93.24 N/mm$^2$ 

Assuming Fixed Connection:
  k1=0.2 k2=0.35 k3=20
  $\lambda_{ef}$=sqrt(k1+k2*$\lambda_{vv}^2$+k3*$\lambda_{\phi}^2$)=0.87
  $\phi$=0.5*[1+$\alpha$($\lambda_{ef}$-0.2)+$\lambda_{ef}^2$]=1.04
  c=($\phi$+sqrt($\phi^2$-$\lambda_{ef}^2$))= 1.62
  f_cd=(fy/$\gamma_{mo}$)/sqrt($\phi$+($\phi^2$-$\lambda^2$)))= 140.42 N/mm$^2$ 

Interpolating values for hinged and fixed connection:
  $\lambda_e$=$\lambda_{eh}$-[($\lambda_{eh}$ - $\lambda_{ef}$)*(0.15/0.35)]= 1.09
  $\phi$=0.5*[1+$\alpha$($\lambda_{e}$-0.2)+$\lambda_{e}^2$]= 1.31
  c=($\phi$+sqrt($\phi^2$-$\lambda_{e}^2$))=2.04
  f_cd=(fy/$\gamma_{mo}$)/sqrt($\phi$+($\phi^2$-$\lambda^2$)))=111.52 N/mm$^2$ 

Design Compressive Force:(f_cd*A)=16.17 kN > 2.02 kN

Check For Slenderness Ratio Limits:- 
  Effective Length of the member=  Le=(Lo*1.0):1.35 m 
  Minimum Radius of Gyration=Rmin:1.27 mm
  Slenderness Ratio of the member= 
  (Le*100)/Rmin:105.72 < 180 

Provide 2 angle 20 x 20 x 4 with weld size 4 mm all 
along all three edges.
\end{verbatim}

\end{document}